# Temperature-dependent thermodynamic properties of $CrNbO_4$ and $CrTaO_4$ by first-principles calculations

Shuang Lin[1], Shun-Li Shang[1], Allison M. Beese[1,2], and Zi-Kui Liu[1]

[1] Department of Materials Science and Engineering, The Pennsylvania State University, University Park, PA 16802, USA

[2] Department of Mechanical Engineering, The Pennsylvania State University, University Park, PA 16802, USA

**Abstract**

In the present work, the density functional theory (DFT) in the generalized-gradient approximation developed by Perdew, Burke, and Ernzerhof (PBE) +U method, i.e., PBE+U, was employed to predict temperature-dependent thermodynamic properties of the rutile-type oxides $CrNbO_4$ and $CrTaO_4$ as well as the binary oxides $Cr_2O_3$, $Nb_2O_5$, and $Ta_2O_5$ via the quasiharmonic phonon approach (QHA). Calculated thermodynamic properties of the binary oxides were benchmarked with experimental data, showing high accuracy except for the negative thermal expansion (NTE) of $Nb_2O_5$, attributed to its polymorphic complexity. By combining the formation energy predicted by DFT with the existing SGTE Substances Database (SSUB5), the $CrNbO_4$ and $CrTaO_4$ are found to be thermodynamic stable up to 1706 K and 1926 K and decompose into $Cr_2O_3$ and $Nb_2O_5$ or $Ta_2O_5$ at those temperatures, respectively. The temperature dependence of linear thermal expansion coefficients for $CrNbO_4$ and $CrTaO_4$ are predicted, and their mean values from 500 K to 2000 K are found to be $6.0\times10^{-6}$/K and $5.04\times10^{-6}$/K, respectively, in agreement with experimental observations in the literature. The gas-phase species and their vapor pressure are calculated, indicating that the formation of $CrTaO_4$ and $CrNbO_4$ reduces chromium volatilization, which is critically important to design enhanced Refractory high entropy alloys (RHEAs) with enhanced oxidation resistance.

## 1. Introduction

Refractory high entropy alloys (RHEAs) are increasingly recognized as promising materials for applications in ultrahigh high-temperature environments, due primarily to their superior mechanical properties at high temperatures [1]. For example, the RHEAs of MoNbTaW and MoNbTaVW demonstrate exceptional yield stresses of 421–506 MPa and 656–735 MPa, respectively, in the 1400–1600°C range with values significantly higher than those of Ni-based superalloys, such as Inconel 718 and Haynes 230, which typically lose strength at such high temperatures [2].

However, significant concerns have been raised regarding oxidation resistances of RHEAs at elevated temperatures. For example, $V_2O_5$ melts at temperatures above 675°C, while $MoO_3$ and $WO_3$ are prone to evaporation at temperatures exceeding 795°C and 1000°C, respectively [3]. While $Cr_2O_3$ has been recognized as a protective oxide in stainless steels like grades 304 and 316 [4,5], as well as in nickel-based superalloys like Inconel alloys [6], its use in refractory metals for ultra-high-temperature applications is limited. This is because $Cr_2O_3$ can react with oxygen to form volatile chromium species, such as CrO and $CrO_2$, which can deplete the protective oxide layer at temperatures above 1000°C [7]. The most critical challenge with RHEAs is the difficulty in forming coherent, self-protective oxide scales on their surfaces, such as $Al_2O_3$, $Y_2O_3$ and $SiO_2$. This limitation arises from the very low solubility of Al, Y or Si in refractory metals (like Nb, Mo, Ta, and W) [8], which are essential for forming these protective oxides. Additionally, incorporating elements like Si into RHEAs to promote the formation of $SiO_2$ can increase the risk of internal oxidation or corrosion attack, as Si can react internally with oxygen, leading to the formation of brittle phases and the degradation of the alloy's mechanical properties [9].

In addition to $Al_2O_3$ and $Cr_2O_3$, recent studies have highlighted the effectiveness of two rutile-type oxides, $CrNbO_4$ and $CrTaO_4$ [10,11], in enhancing oxidation resistance of various alloys. For example, investigations have demonstrated an improved thermal cycling oxidation resistance in Nb-Cr-Si-based alloys. This improvement is linked to the formation of coherent $CrNbO_4$ layers instead of $Nb_2O_5$ [9,11,12]. Additionally, the presence of $CrNbO_4$ at the interface between metal and oxide in the Nb-Si based alloys contributes to a stronger adherence between the oxide scale and the substrate [13]. In the case of the equimolar TaMoCrTiAl RHEAs at 1000°C, the alloy's exceptional oxidation resistance in air is attributed to the formation of $CrTaO_4$ oxide layers. [10]. These layers play a crucial role in significantly impeding oxygen diffusion, thereby enhancing alloys' performance under high-temperature conditions.

The present study aims to predict temperature-dependent thermodynamic properties of $CrNbO_4$ and $CrTaO_4$, including heat capacity (Cp), entropy (S), and linear coefficient thermal expansion (LCTE), in terms of the quasiharmonic phonon approach (QHA) through density functional theory (DFT) based first-principles calculations. Their phase stability and vapor pressures are then investigated by combining with the existing SGTE Substances Database (SSUB5) [14] using Thermo-Calc [15]. Having these thermodynamic properties available is critically important to RHEAs with enhanced oxidation resistance.

## 2. First-principles calculations

2.1 First-principles quasiharmonic phonon approach

DFT-based first-principles calculations provide quantities linked to electronic structure. Helmholtz energy (F) of a given configuration as a function of volume (V) and temperature (T) within QHA is represented by [16–18],

$$F(V,T) = E_c(V) + F_{vib}(V,T) + F_{ele}(V,T) \qquad \text{Eq. 1}$$

where $E_c$ represents the static energy at 0 K without zero-point energy, predicted by DFT directly, $F_{vib}$ the vibrational contribution to Helmholtz energy, and $F_{ele}$ the thermal electronic contribution to Helmholtz energy. The last term $F_{ele}$ is only important for conductors (ignored in the present study of oxides), derived through the integration over the electronic density of states (DOS) with the Fermi-Dirac distribution [16–18].

The present work adopts a four-parameter Birch-Murnaghan equation of state (EOS) to model $E_c$ based on E-V data points from DFT predictions [17],

$$E_c(V) = a + bV^{-2/3} + cV^{-4/3} + dV^{-2} \qquad \text{Eq. 2}$$

where *a, b, c,* and *d* are fitting parameters. The four equilibrium properties under zero external pressure (used in the present work) from this EOS fitting include the equilibrium volume ($V_0$), energy ($E_0$), bulk modulus ($B_0$) and its pressure derivative ($B_0'$).

$F_{vib}$ is obtained from the DFT-based phonon DOS's within QHA [16–18],

$$F_{vib}(V,T) = k_B T \int_0^\infty \ln\,[2sinh\frac{\hbar\omega}{2k_B T}]g(\omega)d(\omega) \qquad \text{Eq. 3}$$

where $k_B$ denotes the Boltzmann constant, $\hbar$ the reduced Planck constant, and $g(\omega)$ the phonon DOS with $\omega$ being frequency.

In the present work, the entropy S was obtained from the Gibbs energy G at constant pressure (P=0 Pa will be used in the present work and G=F):

$$S = -(\frac{\partial G}{\partial T})_P \qquad \text{Eq. 4}$$

Heat capacity at constant pressure $C_p$ was derived from the second derivative of G:

$$C_p = -T(\frac{\partial^2 G}{\partial T^2})_P \qquad \text{Eq. 5}$$

The linear thermal expansion coefficient α at fixed pressure P was calculated as,

$$\alpha = \frac{1}{3V_{0T}}(\frac{\partial V_{0T}}{\partial T})_P \qquad \text{Eq. 6}$$

where $V_{0T}$ is the equilibrium volume at the temperature of interest, determined by the minimization of Helmholtz energy F in the Eq. 1 with respect to volume for P=0 Pa.

## 2.2 Details of first-principles and phonon calculations

All DFT-based first-principles calculations in the present work were performed by the Vienna Ab initio Simulation Package (VASP) [19] using the projector augmented wave (PAW) method [20]. The exchange-correlation functional was described by the generalized gradient approximation developed by Perdew, Burke, and Ernzerhof (PBE) [21]. The energy convergence criterion of electronic self-consistency was chosen as $10^{-6}$ eV per atom in all the calculations. For static calculations at 0 K, we adopted the rotationally invariant PBE+U approach of Liechtenstein et al.

and [22] which have shown a good accuracy in predicting thermodynamic properties of oxides. Integration was performed using the tetrahedron method with Blöchl corrections [23]. The outcome of these calculations is contingent upon the effective on-site Coulomb (U) and exchange (J) parameters, which were set to U = 4.5 eV and J = 1.0 eV for the d orbitals of Cr, used for $Cr_2O_3$ in the literature [24]. Different *k*-point meshes (details in Table 1) together with a 400 eV plane wave cutoff energy were employed based on our convergence tests.

For phonon calculations, we employ the supercell method [25] and the YPHON package [26], with VASP as the computational engine. The supercell sizes used are 80-, 98-, 56-, 24-, and 24-atoms for $Cr_2O_3$, $Nb_2O_5$, $Ta_2O_5$, $CrNbO_4$, and $CrTaO_4$, respectively, along with the corresponding k-point meshes of (4×4×4), (1×1×1), (4×4×2), (6×6×7) and (6×6×7), as detailed in Table 1. For the large supercell of $Nb_2O_5$, a single Γ point 1 × 1× 1 was used as k-point mesh.

The employed unit cells for DFT-based calculations are also listed in Table 1. The unit cell of α-$Cr_2O_3$ (represented by $Cr_2O_3$ in the present work) is rhombohedral and contains 10 atoms in this primitive cell [27]. As shown in Figure 1(a) the spins of the four Cr ions in the unit cell of $Cr_2O_3$ are arranged along the [111] rhombohedral axis in a spin sequence ↑↓↑↓, where symbol "↑" denotes a spin up Cr, and "↓" a spin down Cr. The antiferromagnetic (AFM) configuration with four Cr atoms in the primitive cell results in the ground-state configuration [28–30]. The unit cell of $Cr_2O_3$ exhibits the 0 K lattice parameter of 5.3412 Å and angle of 55.10° [24], which are within 0.2% of experimental values of 5.35 Å and 55.15° [27]. For phonon calculations, $Cr_2O_3$ was transformed into a 2×2×2 supercell, resulting in a structure with 80 atoms, as shown Figure 1(b).

$Nb_2O_5$ can exist in an amorphous state and crystallizes into several types of polymorphs [31]. Among these polymorphs, the monoclinic H-$Nb_2O_5$ (H denoting its high crystallization temperature of 1273 K) [32] has been reported as the most thermodynamically stable phase at 0 K [33]. It is with 14 formula units, and the 98 atoms unit cell is illustrated in Figure 2. The structure of $Ta_2O_5$ was selected based on the orthorhombic structure with the lowest formation energy, as reported in the Materials Project (structure ID: mp-1392387) [34], which is less than 0.01 eV/atom above the convex hull. The lattice parameters of $Ta_2O_5$ (a = 3.84 Å, b = 3.87 Å, and c = 6.50 Å) agree with the experimental values for the orthorhombic $Ta_2O_5$ structure (a = 3.66 Å, b = 3.89 Å, and c = 6.20 Å) [35].

The structures of $CrNbO_4$ and $CrTaO_4$ are based on the prototype of the rutile phase in $TiO_2$, where Cr and Nb (or Ta) are octahedrally coordinated, each surrounded by six oxygen atoms [34], also in accordance with experimental observations [36]. Figure 3 and Figure 4 depict the structure of CrNbO4 and CrTaO4, and Table 3 shows the Wyckoff sites of the CrNbO4 and CrTaO4. respectively. The four Nb (or Ta) atoms occupy the Wyckoff positions 4a, four Cr atoms occupy 4a positions, and the 16 O atoms occupy the 8b positions.

## 3. Results and discussion

### 3.1 Equation of state of binary oxides: $Cr_2O_3$, $Nb_2O_5$ and $Ta_2O_5$

Table 2 shows the predicted equilibrium properties of binary oxides, i.e., $V_0$, $E_0$, $B_0$, and $B'$, from EOS fitting at 0 K, together with experimental bulk modulus in the literature.

For $Cr_2O_3$, PBE+U predicts $B_0$ = 199 GPa, aligning well with the experimentally measured value of 200 GPa reported by Jõgiaas et al. [37]. However, it is lower than the values reported by Rossi and Lawrence. (232 GPa) [38] and Rekhi et al. (240 GPa) [39]. On the other hand, the PBE+U prediction is lower than the LDA+U result, which gives $B_0$ = 238 GPa. The difference between PBE+U and LDA+U stems from that LDA underestimates the lattice parameters compared to PBE [40].

For H-$Nb_2O_5$, the PBE prediction is compared with the DFT calculations using PBEsol in the literature [33]. While PBE predicts the same $E_0$ value as PBEsol, it gives an equilibrium volume ($V_0$) of 14.70 Å$^3$/atom, larger than the 13.86 Å$^3$/atom predicted by PBEsol. This difference in equilibrium volume leads to a lower $B_0$ of 129 GPa predicted by PBE, which is 29 GPa lower than the 158 GPa predicted by PBEsol. This difference arises because PBEsol was specifically designed to improve the description of solids and solid surfaces near equilibrium, leading to underestimated lattice constants compared to PBE [41,42]. For $Ta_2O_5$, PBE predicts a $B_0$ of 186 GPa, which is higher than the experimentally measured value of 138 GPa [43]. Compared to the volume at 0 K, the volume measured at ambient temperature is larger due to thermal expansion, which reduces atomic forces and consequently lowers $B_0$.

## 3.2 Thermodynamic properties of binary oxides: $Cr_2O_3$, $Nb_2O_5$ and $Ta_2O_5$

Figure 5 depicts the predicted entropies of these three binary oxides. The entropy of $Cr_2O_3$ in the present study is 2 J/mol·K lower than that in the SSUB5 database at 300 K and up to 10 J/mol·K lower at higher temperatures. The predicted entropy values for $Nb_2O_5$ agree well with the results from the SSUB5 database and those by Jacob et al. [44] from 298.15 K to 1700 K. The predicted entropy for $Ta_2O_5$ aligns closely with the SSUB5 database and Jacob et al. [45] until 1593 K, beyond which phonon predictions give a lower entropy due to the phase transition from β to α.

In Figure 6, Cp of $Cr_2O_3$ measured by Anderson [46] and Gurevich et al. [47] exhibits a lambda-type shape near the Néel temperature 307 K, which is not considered in the present study. This could be addressed by the zentropy approach [48–50] though beyond the scope of the current work. The predicted Cp of $Cr_2O_3$ show minimal differences when compared to phonon calculations using the LDA+U approach [24], in very good agreement with SSUB5 and the high-temperature measurements by Ziemeniak et al. [51] with a standard deviation ±3% at temperatures below 1673 K and ±5% between 1673 and 1873 K. The predicted Cp values of $Nb_2O_5$ agree well with SSUB5 and the reassessment by Jacob et al. [44] up to 800 K, with a slight underestimation observed thereafter, up to 2.0 J/mol·K. Similarly, the Cp of $Ta_2O_5$ is in agreement with the measurements made by Orr [52] and SSUB5 up to 700 K, but underestimated at higher temperature.

Figure 7 plots the calculated LCTE of the three binary oxides with experimental data superimposed. For $Cr_2O_3$, the present work predicts higher LCTE values than those by Wang et al. [24] with the largest difference being $0.5\times10^{-6}$/K, due to the LDA potential by Wang et al. [24] which underestimates lattice parameters, resulting in stiffer interatomic force constants [53]. Furthermore, the measured LCTE values of $Cr_2O_3$ are quite scattered, ranging from $5.7\times10^{-6}$/K to $10.75\times10^{-6}$/K [24,54–56] .

It is noted that QHA predicts a negative thermal expansion (NTE) for H-$Nb_2O_5$ from 0 to 250 K, which is supported by the literature evidencing NTE at low temperatures. Manning et al. [57] investigated LCTE of the $Nb_2O_5$ during heating and cooling cycles ranging from 273 K to 1473 K and observed NTE during heating and cooling below 445 K and 485 K, respectively. Manning et al. [57] also reported that the LCTE values along the a-, b-, and c-axis are $5.3\times10^{-6}$/K, 0, and $5.9\times10^{-6}$/K, respectively. Durbin and Harman [58] reported a low thermal expansion between -$1.2\times10^{-6}$/K to $2.0\times10^{-6}$/K from 298K to 1173K. Douglass [59] measured the LCTE of $Nb_2O_5$ from 473K to 1273K, reporting values ranging from $1.0 \times 10^{-6}$/K to $2.1 \times 10^{-6}$/K, with the LCTE remaining positive throughout. The significant variation of experimental measurement in the LCTE of $Nb_2O_5$ can be attributed to its complex polymorphism, with over 15 reported structures [60,61]. Brauer [62] identified three primary crystalline forms of $Nb_2O_5$, i.e., amorphous to 500°C, γ to 1000°C, β to 1100°C, and α-phase above. This complex polymorphic transformation into different structures supports and contributes to the observed variability in $Nb_2O_5$'s thermal expansion data. The difference between our predicted LCTE and the experimentally measured LCTE of $Nb_2O_5$, could potentially be resolved by considering both ground-state and metastable non-ground-state configurations, as described by the zentropy theory [48–50], though beyond the scope of the present study.

For $Ta_2O_5$ thin films, Yoon et al.[63] and Tien et al. [64] reported a mean LCTE of $4.68\times10^{-6}$/K from 323K to 473K, and $3.6\times10^{-6}$/K between 298K to 328K, respectively. Wu et al. [65] reported a LCTE of $6.72\times10^{-6}$/K from room temperature to 1273K for $Ta_2O_5$. The predicted LCTE values for $Ta_2O_5$ closely align with these experimental results with a deviation of about $1\times10^{-6}$/K.

3.3 Ground state configurations and thermodynamic properties of $CrNbO_4$ and $CrTaO_4$

Due to the magnetic nature of Cr, we examine the energetics of $CrNbO_4$ and $CrTaO_4$ using four independent 24-atom spin-polarized configurations as determined by symmetry analysis with the ATAT [66]. The magnetic configurations of Cr are detailed in Table 3 with Cr1 through Cr4 corresponding to the positions of Cr in Figure 3 and Figure 4 for CrNbO4 and CrTaO4, respectively, with "↑" and "↓" for spin up and down of Cr. The relaxed energies indicate that the spin configurations of Cr do not substantially change the energy of $CrNbO_4$, evidenced by the small energy differences among these four configurations, which are within 3 meV/atom. The configuration with the lowest energy is designated as str1, displaying a ferromagnetic configuration of Cr in $CrNbO_4$.

In $CrTaO_4$, ferromagnetic configuration str1 also gives the lowest energy among four spin configurations. The energies of str1, str2, and str4 are very close, within 1meV/atom difference. A significant energy difference of 7 meV/atom is observed between the energies of str1and str3, indicating the CrTaO4 str3 antiferromagnetic structure is the least stable magnetic configuration.

Employing the same QHA used for binary oxides, entropy, heat capacity and LCTE of $CrNbO_4$ and $CrTaO_4$ are predicted as shown in Figure 8. The entropy values of $CrNbO_4$ and $CrTaO_4$ are close to each other up to 1000 K. At higher temperature, the entropy of $CrNbO_4$ becomes higher than that of $CrTaO_4$ by around 1 J/mol·K at 1000 K. Figure 8 presents the predicted Cp values for both oxides, showing that the Cp of $CrNbO_4$ is slightly higher than that of $CrTaO_4$.

The Gibbs energy of formation ($\Delta G_f$) was calculated based on the obtained entropy and enthalpy data (detailed enthalpy values for each oxide are provided in Table S1 in supplementary file). Table 5 presents $\Delta G_f$ for $CrNbO_4$ and $CrTaO_4$ as a function of temperature with respect to their respective binary oxides, $Cr_2O_3$ and $Ta_2O_5$ or $Nb_2O_5$, derived from DFT-based QHA. $\Delta G_f$ for $CrNbO_4$ and $CrTaO_4$ are both negative (-6993 J/mol-formula and -1538 J/mol-formula, respectively), indicating that they are thermodynamically stable at 0K. As the temperature increases to 2000 K, as shown in the Figure 10, the $\Delta G_f$ of $CrNbO_4$ continues to decrease to -9640 J/mol. In contrast, the $\Delta G_f$ of $CrTaO_4$ varies less with temperature, increasing to -1235 J/mol from 0 to 1000 K and then decreases to -1700 J/mol at higher temperatures. By combining $\Delta G_f$ with SSUB5 [14], $CrNbO_4$ and $CrTaO_4$ are found to be thermodynamic stable up to 1706 K and 1926 K and decompose into $Cr_2O_3$ and $Nb_2O_5$ or $Ta_2O_5$ at those temperatures, respectively. The combined database will also be utilized in Section 4 in the context of calculating the vapor pressures and gaseous species associated with $CrNbO_4$ and $CrTaO_4$.

Hong's machine learning model [67] predicts the congruent melting temperatures to be $1776 \pm 57$ K for $CrNbO_4$ and $2239 \pm 65$ K for $CrTaO_4$. The experimentally measured melting temperature of $CrTaO_4$, prepared by hot-press sintering of powders synthesized through solid-state reaction, was recently reported to be 2103 ± 20 K [68]. The decomposition temperatures predicted from thermodynamic database are lower than the machine learning-predicted or experimental congruent melting points, due to the absence of liquid phase descriptions for $CrNbO_4$ and $CrTaO_4$ in the thermodynamic database.

Table 6 presents the predicted LCTEs of $CrNbO_4$ and $CrTaO_4$ from this work, alongside previously DFT-predicted and experimental data. Tabero [69] used high-temperature X-ray diffractometry to determine the LCTE of $CrNbO_4$, reporting values of $5\times10^{-6}$/K, $5\times10^{-6}$/K, and $8\times10^{-6}$/K along the a-, b-, and c-axes, respectively, from 298 to 1073 K. Our predicted mean LCTE values of $CrNbO_4$ agrees well with these results and fall within this range. Hao et al. [8] predicted a LCTE of $13.6\times10^{-6}$/K for $CrTaO_4$ over the temperature range of 500–2000 K by first-principles calculations, significantly higher than our predictions, likely due to differences in the magnetic settings of Cr in $CrTaO_4$. In their study, spin-orbit coupling was employed, whereas we specify spin polarization as ferromagnetic configuration. Our prediction of LCTE is slightly lower and but agrees well with experimentally measured LCTE of $CrTaO_4$ reported by Zhang et al. [68], which were $\alpha_a=(5.68\pm0.10)\times10^{-6}$/K and $\alpha_c=(7.81\pm0.11)\times10^{-6}$/K, yielding an average thermal expansion coefficient of $(6.39\pm0.11)\times10^{-6}$/K from 560 to 1473 K.

## 4. Vapor pressure and gaseous species of $CrNbO_4$ and $CrTaO_4$

The vapor pressure and gaseous species of $CrNbO_4$ and $CrTaO_4$ were calculated by combining the SSUB5 database [14], with the Gibbs energy of $CrNbO_4$ and $CrTaO_4$, respectively. In the calculations, the ratio of the number of moles of Cr or Nb with respect to Ta was set to 1:1, and three oxygen activities were considered, i.e., $10^{-3}$, $10^{-5}$, and $10^{-7}$ with respect to pure oxygen gas at the temperature of calculations, i.e., $ACR(O_2)=P(O_2)/P$ in Thermo-Calc [15] with $P(O_2)$ and P being the $O_2$ partial pressure and total pressure, respectively. The amount of gas phase was fixed at zero in all simulations.

Figure 10 presents the total vapor pressure of $CrNbO_4$ (a) and $CrTaO_4$ (b) calculated at three oxygen activities. For both oxides, vapor pressure keep constant at low temperature, and then increases significantly beyond 2000 K. At low temperatures, where $O_2$ is the dominant gas-phase species, the total vapor pressure is primarily governed by the oxygen activity, with higher $ACR(O_2)$ values resulting in higher total pressure. At high temperatures, Cr-containing species such as $Cr_1O_2$ and $Cr_1O_2$ begin to evaporate. Under lower $ACR(O_2)$ conditions, the gas phase contains a higher proportion of Cr-containing species leads to a higher total vapor pressure.

Figure 11 shows the calculated partial pressures of individual gaseous species for $CrNbO_4$ (a) and $CrTaO_4$ (b) with $ACR(O_2) = 10^{-3}$. At lower temperatures, $O_2$ is the dominant gaseous species. As the temperature increases, the partial pressures of chromium oxides ($Cr_1O_2$ and $Cr_1O_2$) rise, but remain lower than $P(O_2)$. Low partial pressures of Cr and Cr-containing gaseous species lead to reduced chromium loss, thereby helping to preserve the integrity of the protective oxide scale. As a result, the long-term oxidation resistance and thermal stability of $CrNbO_4$, and $CrTaO_4$ are significantly enhanced, making these materials well-suited for high-temperature applications where volatilization is a critical concern.

Among all gaseous species, Cr-containing oxides exhibit significantly higher vapor pressures than Nb- or Ta-containing species. As shown in Figure 11, Nb species (e.g., Nb, NbO) and Ta species (e.g., Ta, TaO) display much lower vapor pressures, indicating greater thermodynamic stability under the same conditions.

Two additional cases with lower oxygen activities, $ACR(O_2)= 10^{-5}$ and $10^{-7}$, were also evaluated, with detailed figures (Figure S1 to Figure S4) provided in the supplementary material. The results show that at low temperatures, $O_2$ remains the dominant gaseous species. However, at higher temperatures approximately 2000 K for $ACR(O_2) = 1e^{-5}$ and 1800 K for $ACR(O_2) = 1e^{-7}$, the partial pressures of $Cr_1O_2$ and $Cr_1O_3$ surpass that of $O_2$. This behavior is expected under low oxygen activity conditions, where the formation of volatile Cr-containing oxides, such as $CrO_2$ and $CrO_3$, is thermodynamically favored and evaporation begins. Despite the increase in Cr-oxide vapor pressure, the total vapor pressure remains relatively low, even at elevated temperatures such as 2000 K.

## 5. Conclusions

In the present study, the PBE+U approach is employed to predict the thermodynamic properties of $Cr_2O_3$, $Nb_2O_5$, and $Ta_2O_5$, including entropy, heat capacity, and linear thermal expansion coefficient as a function of temperature in comparison with experimental data in the literature. Our predictions closely matched experimental findings for all assessed thermodynamic properties, except for the NTE for $Nb_2O_5$. The benchmarked approach is then used to investigate thermodynamic properties of two rutile-type oxides, $CrNbO_4$ and $CrTaO_4$. It is predicted that from 500 K to 2000 K, the mean LCTEs of $CrNbO_4$ and $CrTaO_4$ were $6.00\times10^{-6}$/K and $5.04\times10^{-6}$/K, respectively, showing reasonable agreement with available experimental data in the literature. The vapor pressure of gas-phase species supports the formation of $CrTaO_4$ and $CrNbO_4$ at $ACR(O_2) = 10^{-3}$, which enhances oxidation resistance by reducing chromium volatilization in the gas phase, a critical property for achieving high-temperature stability in RHEA applications.

## 6. Acknowledgements

The present work is based upon work supported by the Department of Energy/Advanced Research Projects Agency - Energy (ARPA-E) under award No DE-AR0001435. First-principles calculations were performed partially on the Roar Collab supercomputer at the Pennsylvania State University's Institute for Computational and Data Sciences (ICDS), and partially on the resources of the Extreme Science and Engineering Discovery Environment (XSEDE) supported by NSF with Grant No. ACI-1548562.

## 7. Figures

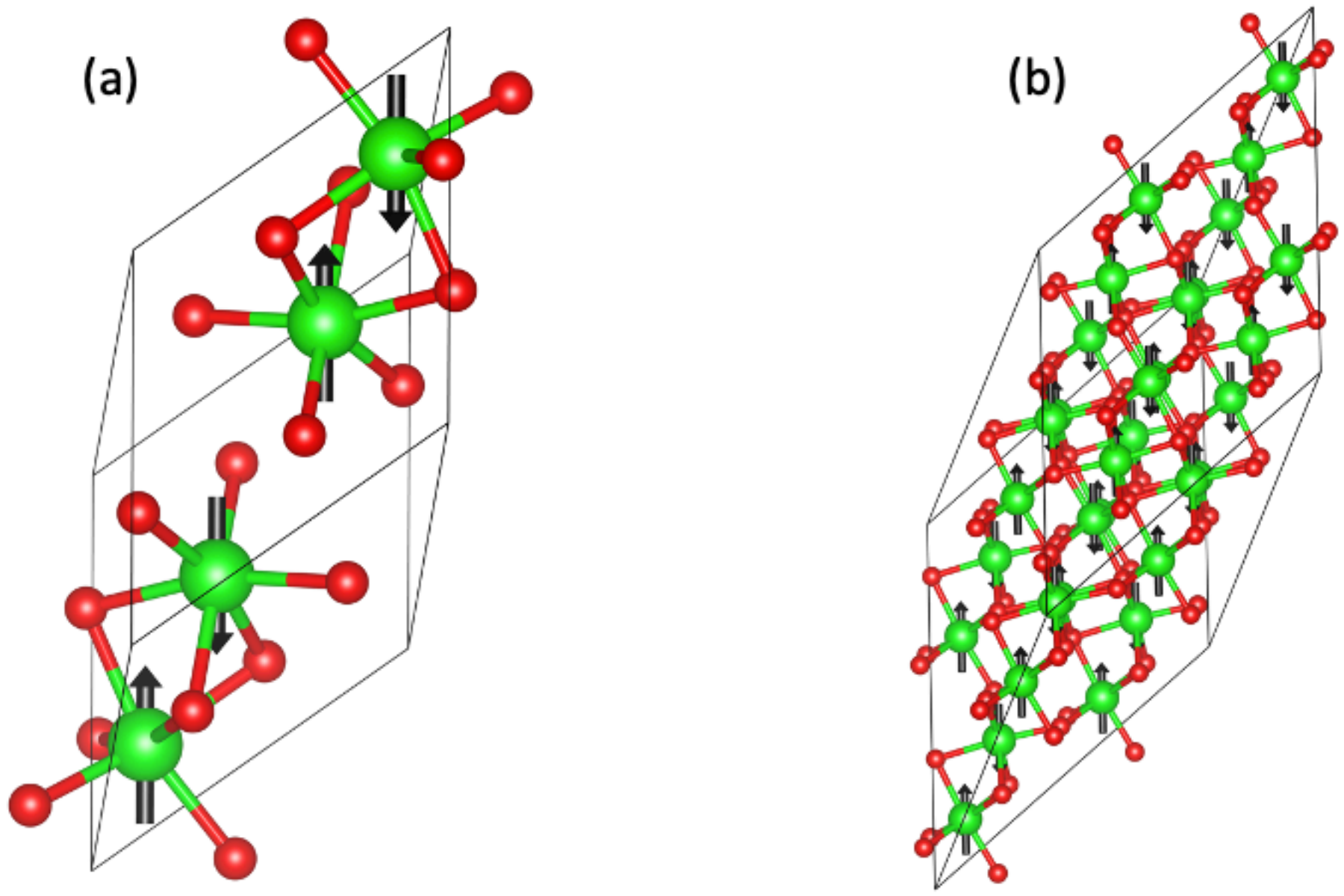

*Figure 1.(a) Unit cell of antiferromagnetic structure of $Cr_2O_3$ where green spheres represent Cr atoms and red spheres represent O atoms. The spins of the four Cr ions within the unit cell are aligned along the [111] rhombohedral direction, with "↑" indicating spin-up Cr and "↓" indicating spin-down Cr. (b) The 2×2×2 supercell of Cr2O3 containing 80 atoms in total, used for performing the phonon calculations.*

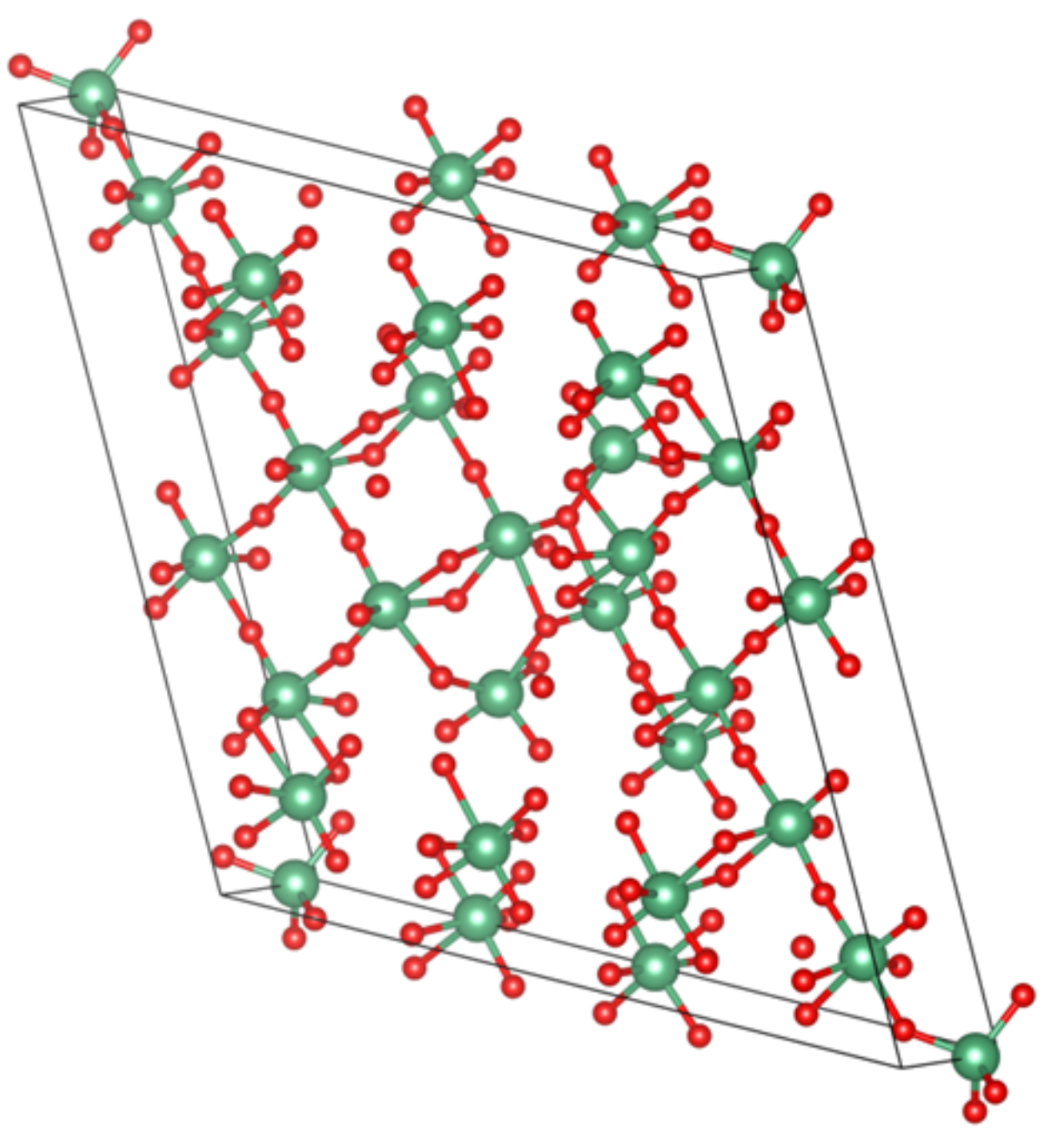

*Figure 2. Atomic structure of H-Nb2O5 with green spheres representing Nb atoms and red spheres representing O atoms. This structure has been reported as thermodynamically the most stable phase at 0 K [33].*

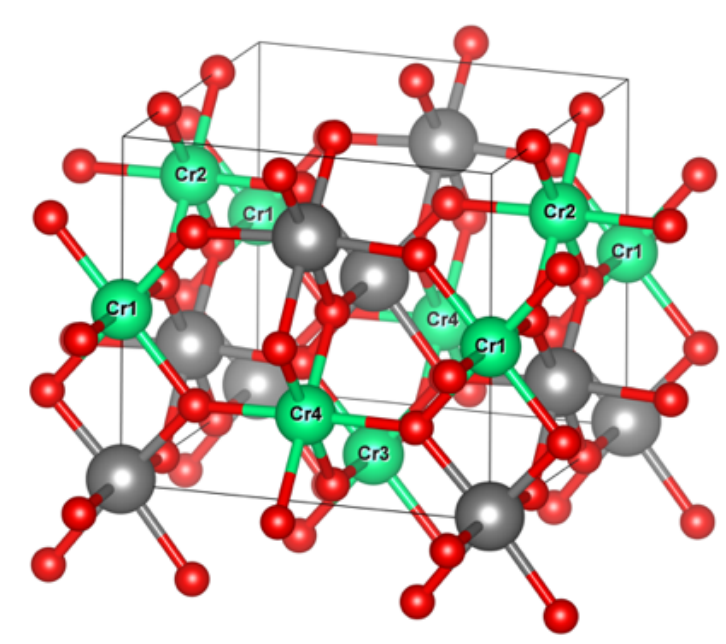

*Figure 3. Crystal structure of $CrNbO_4$: Both Cr and Nb atoms are octahedrally coordinated by six oxygen atoms. Identical numbers following Cr indicate crystallographically equivalent positions.*

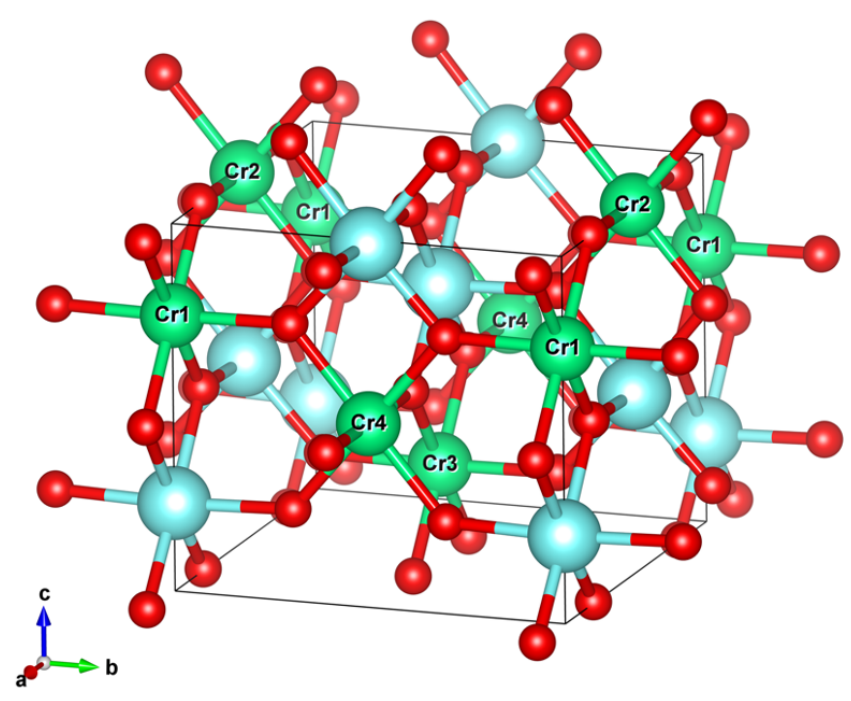

*Figure 4. Crystal structure of $CrTaO_4$: Both Cr and Ta atoms are octahedrally coordinated by six oxygen atoms. Identical numbers following Cr indicate crystallographically equivalent positions.*

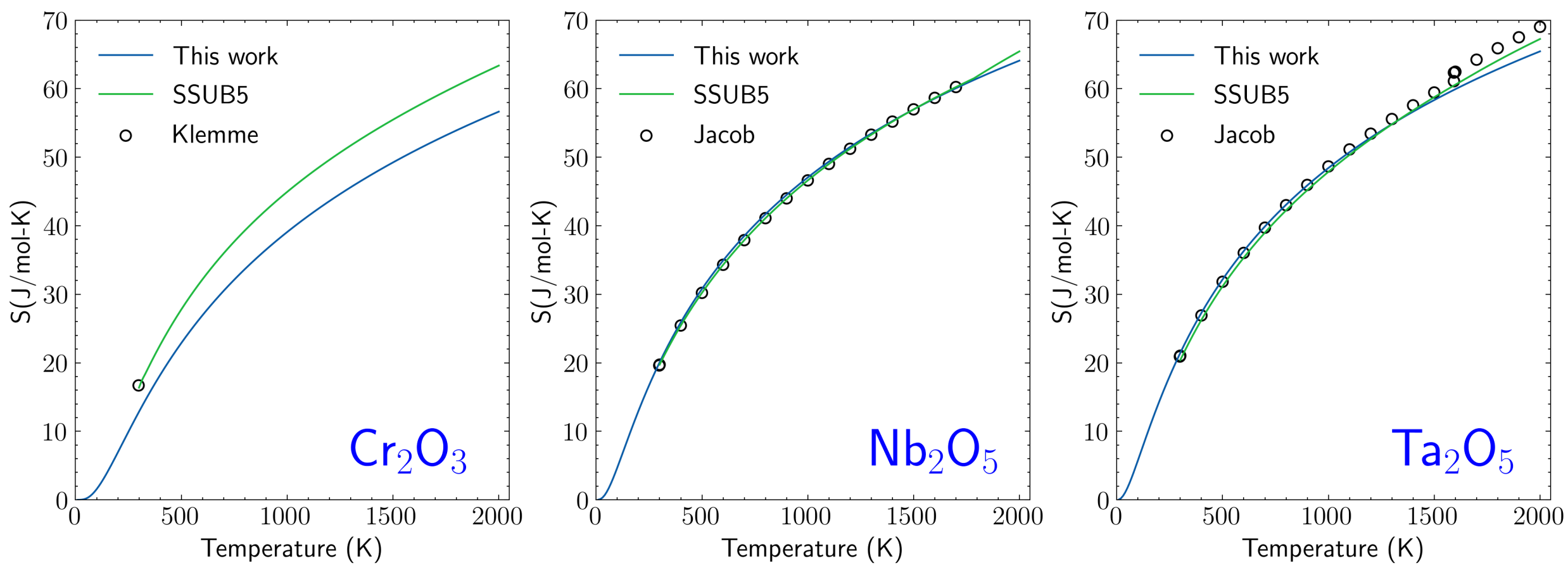

*Figure 5. Entropy of $Cr_2O_3$, $Nb_2O_5$ and $Ta_2O_5$ using QHA method, with comparison to the SSUB5 database [14] and experimental data for $Cr_2O_3$ [70] , $Nb_2O_5$ [71] and $Ta_2O_5$ [45].*

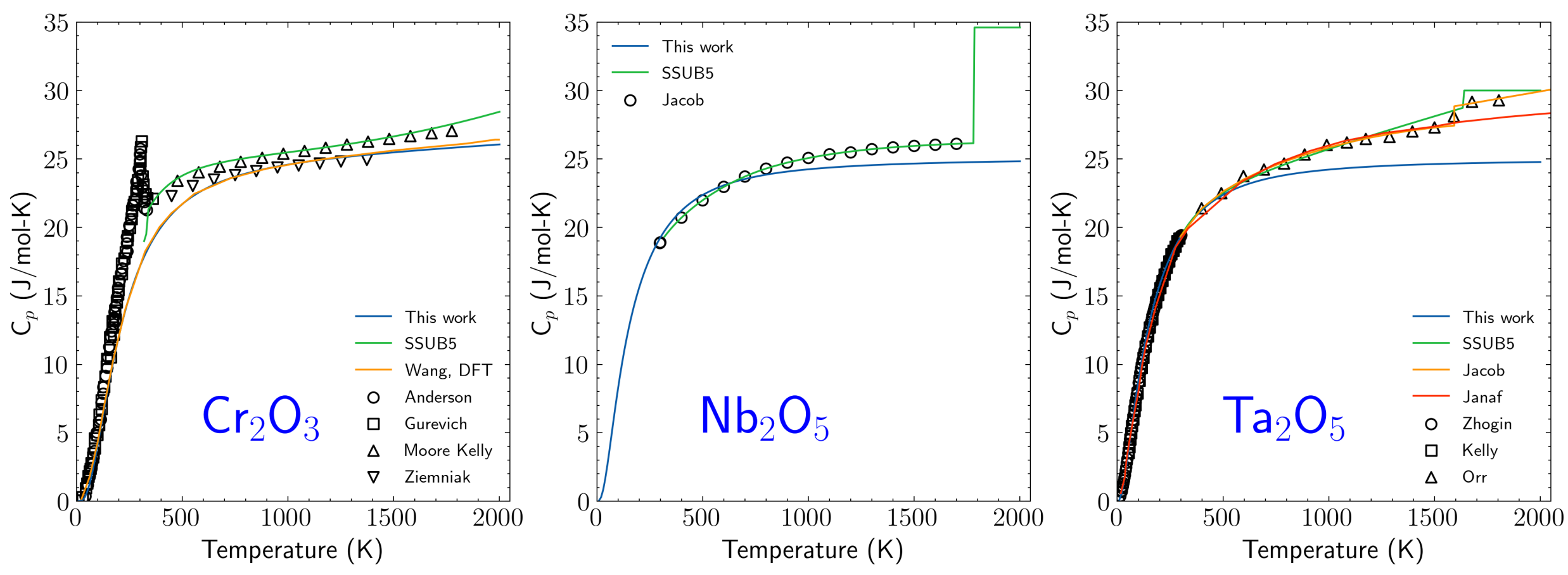

*Figure 6. Heat capacity of $Cr_2O_3$, $Nb_2O_5$ and $Ta_2O_5$ using QHA method, with comparison to SSUB5 database [14], DFT results [24] and experimental data for $Cr_2O_3$ [46,47,51,72], as well as experimental data for $Nb_2O_5$ [71] and $Ta_2O_5$ [45,52,72–74].*

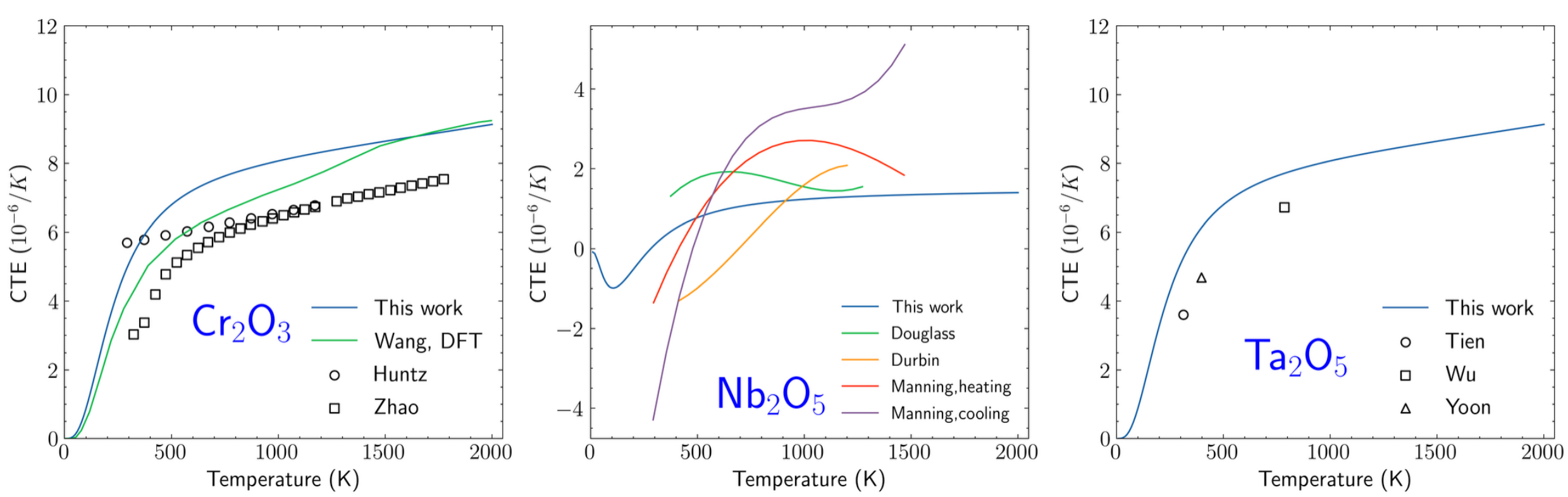

*Figure 7. LCTE of $Cr_2O_3$, $Nb_2O_5$ and $Ta_2O_5$, with comparison to DFT predictions [24] and experimental data for $Cr_2O_3$ [55,56], as well as experimental findings for $Nb_2O_5$ [57–59]and $Ta_2O_5$ [63–65].*

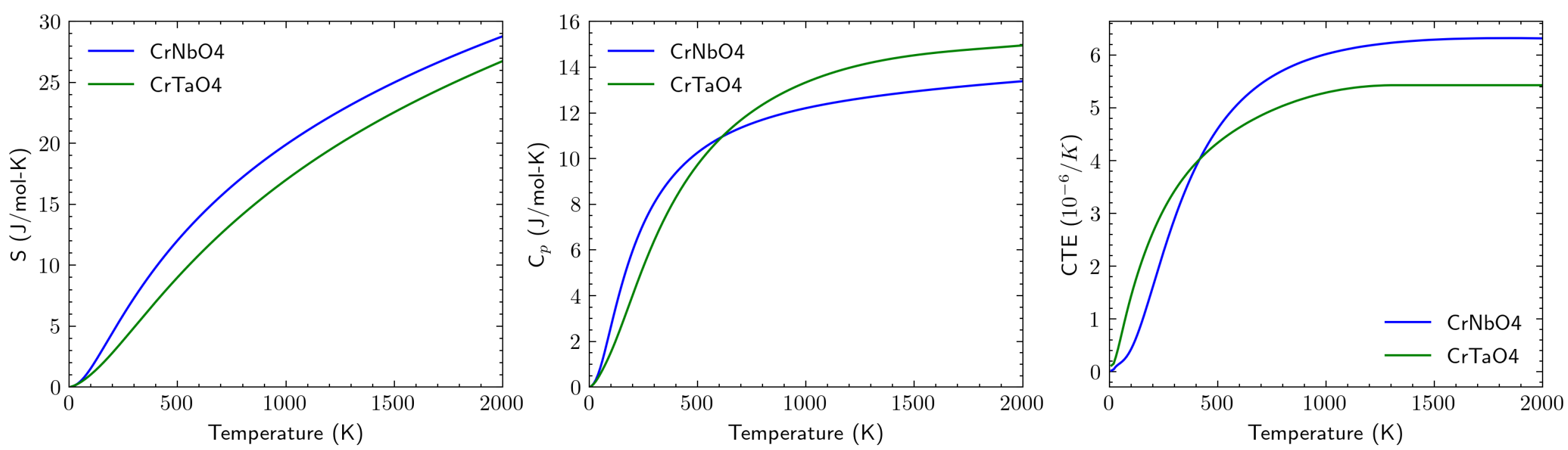

*Figure 8. Thermodynamic properties, i.e., entropy (S), heat capacity (Cp) and linear thermal expansion coefficient (LCTE), prediction of CrNbO4 and CrTaO4 by this work.*

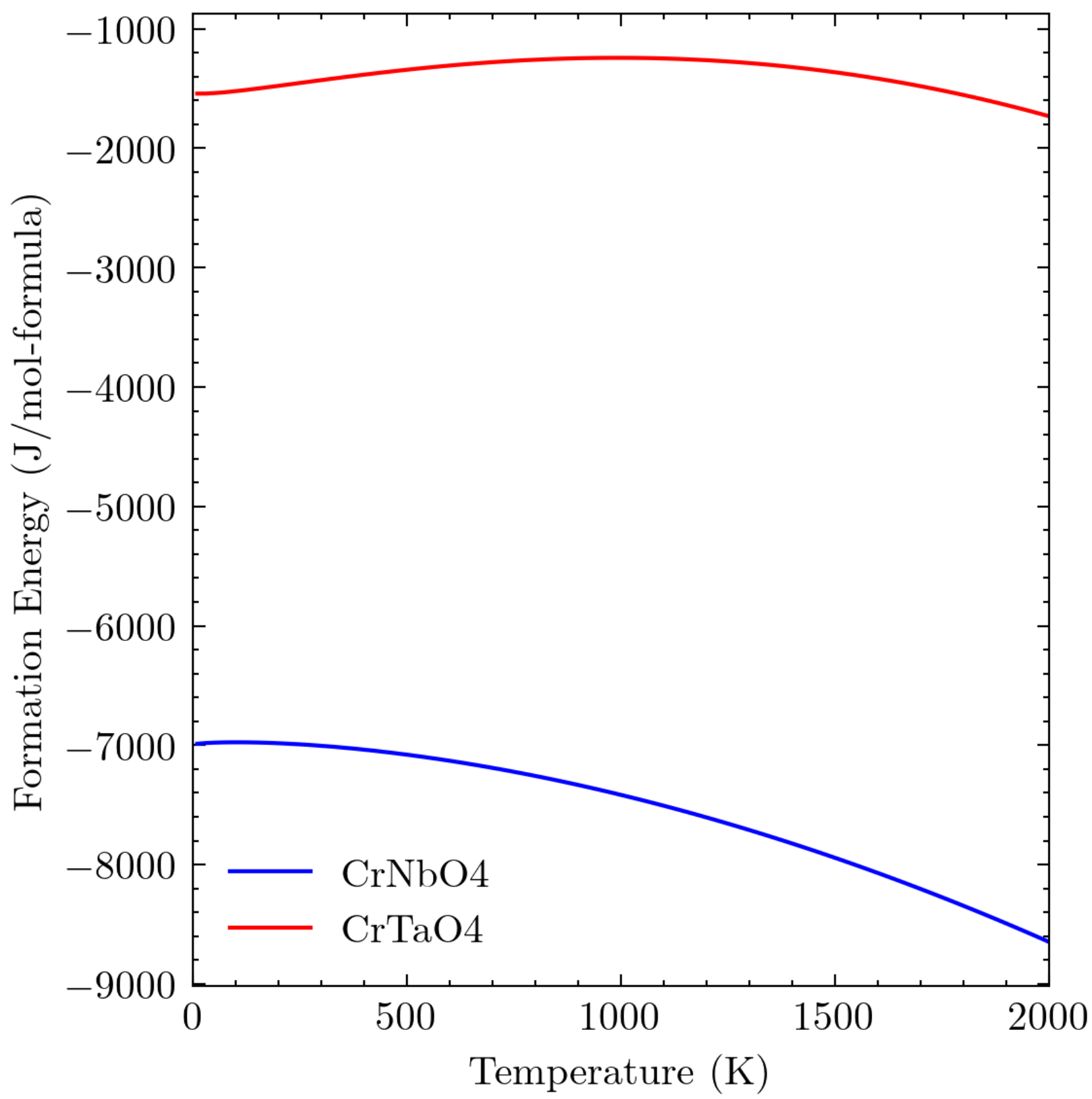

*Figure 9. Formation energy ($\Delta G_f$) of the $CrNbO_4$ and $CrTaO_4$ as functions of temperature with respect to their respective binary oxides, $Cr_2O_3$ and $Ta_2O_5$ or $Nb_2O_5$.*

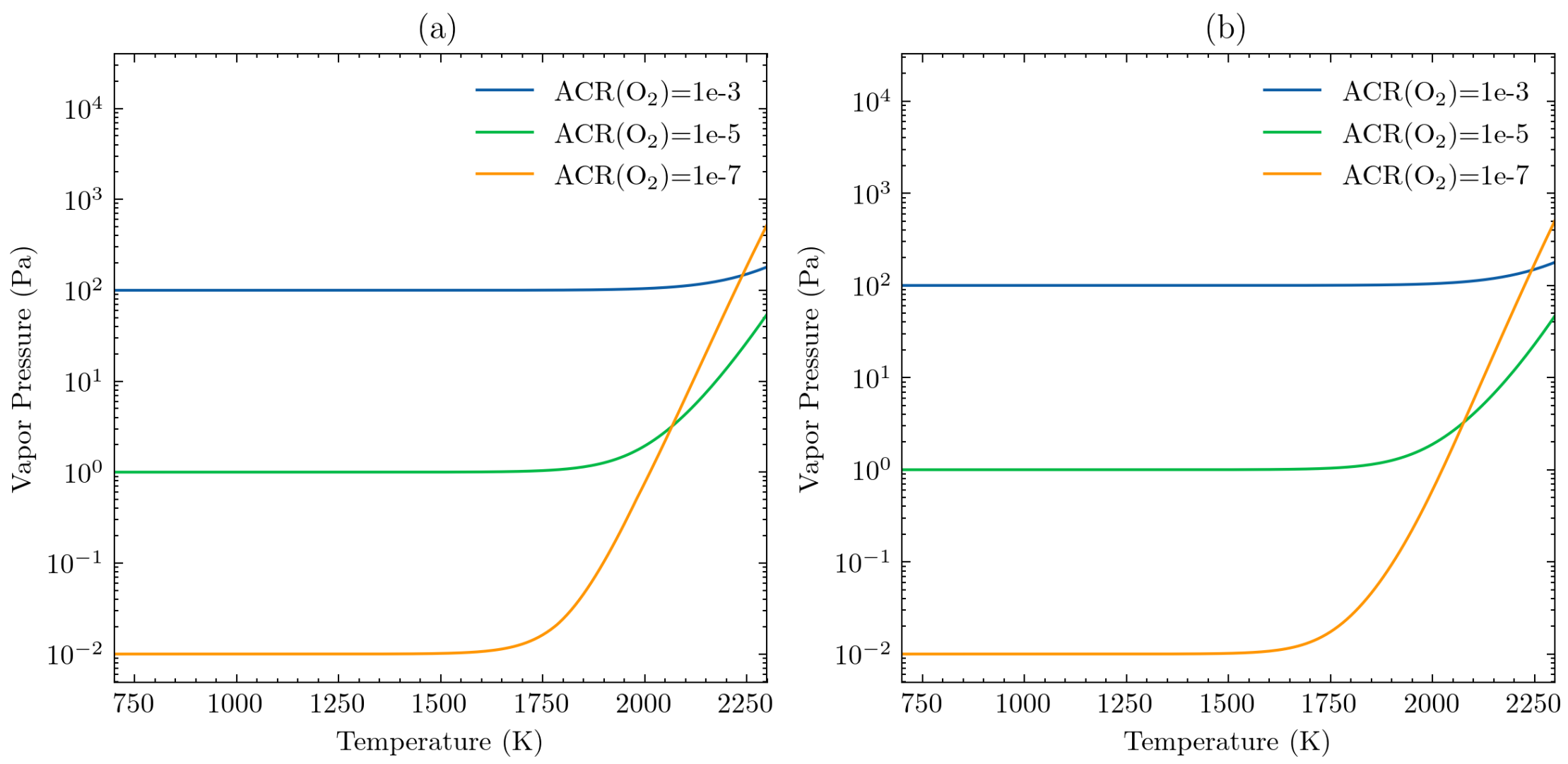

*Figure 10. Temperature-dependent total vapor pressure in (a) $CrNbO_4$ and (b) $CrTaO_4$ under different oxygen activities ($ACR(O_2)$ = $10^{-3}$, $10^{-5}$, and $10^{-7}$).*

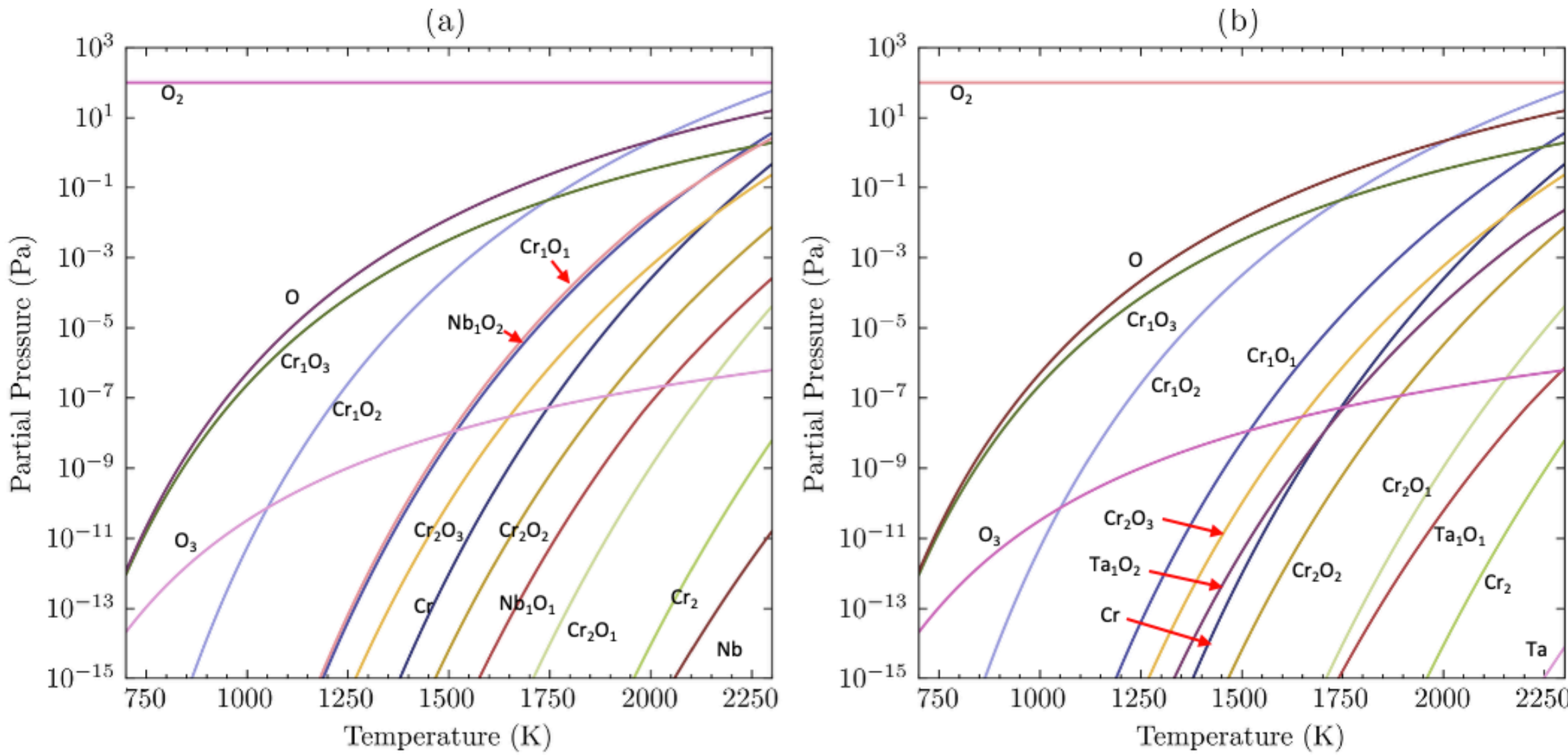

*Figure 11. Temperature-dependent partial pressures of individual gas species for (a) $CrNbO_4$ and (b) $CrTaO_4$ at ACR($O_2$) = $10^{-3}$. While the partial pressures of Cr-containing species ($Cr_1O_3$ and $Cr_1O_2$) increase significantly as the temperature approaches 2000 K, $O_2$ remains the dominant gaseous species across the entire temperature range.*

## 8. Tables

*Table 1. Setting details for DFT-based calculations.*

| Oxides | Space group | Atoms in crystallographic cell | k-mesh electron | Atoms in supercell for phonon | k-mesh phonon | Ref |
|---|---|---|---|---|---|---|
| $Cr_2O_3$ | $R\bar{3}c$ | 10 | 8×8×8 | 80 | 4×4×4 | [24] |
| $Nb_2O_5$ | C12/m1 | 98 | 1×6×1 | 98 | 1×1×1 | [33] |
| $Ta_2O_5$ | Pmmn | 14 | 8×8×2 | 56 | 4×4×2 | mp-1539317 |
| $CrNbO_4$ | $I4_1md$ | 24 | 6×6×7 | 24 | 6×6×7 | mp-758053 |
| $CrTaO_4$ | $I4_1md$ | 24 | 6×6×7 | 24 | 6×6×7 | mp-753467 |

*Table 2. Predicted equilibrium properties of volume $V_0$, energy $E_0$, bulk modulus $B_0$, and first derivative of bulk modulus with respect to pressure $B'$, for each binary oxide from EOS fitting at 0 K. Experimental data and First-principles data are also listed for comparison.*

| Oxides | Method | $V_0$(Å$^3$/atom) | $E_0$ (eV/atom) | $B_0$ (GPa) | B' |
|---|---|---|---|---|---|
| $Cr_2O_3$ | PBE+U | 9.20 | -6.91 | 199 | 6.3 |
| | LDA+U [24] | | | 238 | |
| | Taivo Jõgiaas et al. [37] | | | 200 | |
| | L.R. Rossi, and W.G. Lawrence. [38] | | | 232 | |
| | Sandeep Rekhi et al. [39] | | | 240 | |
| H-$Nb_2O_5$ | PBE | 14.70 | -8.97 | 129 | 2.5 |
| | PBEsol [33] | 13.86 | -8.97 | 158 | |
| $Ta_2O_5$ | PBE | 14.19 | -9.77 | 186 | 6.9 |
| | Matthew C Brennan et al [43] | | | 138 | |

*Table 3. Wyckoff positions of CrNbO4 and CrTaO4*

| | Wyckoff | Element | x | y | z |
|---|---|---|---|---|---|
| CrNbO4 | 4a | Nb | 0.5 | 0 | 0.23691 |
| | 4a | Cr | 0 | 0.5 | 0.248334 |
| | 8b | O | 0.305435 | 0.5 | 0.754222 |
| | 8b | O | 0.29969 | 0.5 | 0.253154 |
| | | | | | |
| CrTaO4 | 4a | Ta | 0.5 | 0.5 | 0.260451 |
| | 4a | Cr | 0 | 0 | 0.250214 |
| | 8b | O | 0.804275 | 0.5 | 0.249535 |
| | 8b | O | 0.801215 | 0.5 | 0.745109 |

*Table 4. Equilibrium volume $V_0$ and energy $E_0$ of in CrNbO4 and CrTaO4 with different Cr spin configurations.*

| | | Cr1 | Cr2 | Cr3 | Cr4 | $V_0$(Å$^3$/atom) | $E_0$(eV/atom) |
|---|---|---|---|---|---|---|---|
| CrNbO4 | Str1 | ↑ | ↑ | ↑ | ↑ | 10.800 | -8.184 |
| | Str2 | ↑ | ↓ | ↑ | ↑ | 10.775 | -8.181 |
| | Str3 | ↓ | ↓ | ↑ | ↑ | 10.800 | -8.184 |
| | Str4 | ↑ | ↓ | ↑ | ↓ | 10.809 | -8.182 |
| CrTaO4 | Str1 | ↑ | ↑ | ↑ | ↑ | 11.031 | -8.555 |
| | Str2 | ↓ | ↑ | ↑ | ↑ | 11.008 | -8.554 |
| | Str3 | ↓ | ↓ | ↑ | ↑ | 10.902 | -8.548 |
| | Str4 | ↓ | ↑ | ↓ | ↑ | 11.059 | -8.555 |

*Table 5. The formation energy ($\Delta G_f$) of $CrNbO_4$ and $CrTaO_4$ as a function of temperature, using their binary oxides as reference states.*

| Equation | $\Delta G_f$ (J/mol) |
| --- | --- |
| $0.5Cr_2O_3+0.5Nb_2O_5=CrNbO_4$ | -6993+0.8*T-0.1*T*ln(T)+3.1e-4*T**2 |
| $0.5Cr_2O_3+0.5Ta_2O_5=CrTaO_4$ | -1538-1.1*T+0.29*T*ln(T)-5.9e-4*T**2 |

*Table 6. Predicted LCTE of $CrNbO_4$ and $CrTaO_4$ compared with DFT and experimental data*

| | Temperature (K) | LCTE ($10^{-6}$ $K^{-1}$) | Reference |
|---|---|---|---|
| CrNbO4 | 500 to 2000 | 6.0 | This work |
| | 298 to 1073 | $\alpha_a$ = 5.0<br>$\alpha_b$ =5.0<br>$\alpha_c$ =8.0 | [69] |
| CrTaO4 | 500 to 2000 | 5.04 | This work |
| | 500 to 2000 | 13.4 | [8] |
| | 560 to 1473 | $\alpha_a$ = (5.68±0.10)<br>$\alpha_c$ = (7.81±0.11)<br>$\alpha_{average}$ = (6.39±0.11) | [68] |